\documentclass[a4paper]{jpconf}
\usepackage{graphicx}
\usepackage{amsmath}

\begin{document}
\title{Interpretation of strange hadron production at LHC}

\author{M~Petr\'{a}\v{n}$^1$, J~Letessier$^2$, V~Petr\'{a}\v{c}ek$^3$ and J~Rafelski$^1$}

\address{$^1$ Department of Physics, University of Arizona, Tucson, AZ 85719, USA}
\address{$^2$ Laboratoire de Physique Th\'{e}orique et Hautes Energies, Universit\'{e} Paris 6, Paris 75005, France}
\address{$^3$ Czech Technical University in Prague, Brehova 7, 115 19 Praha 1, Czech Republic}


\begin{abstract}
We extend the SHM analysis of hadron production results showing here consistency with the increased experimental data set, stability of the fit with regard to inclusion of finite resonance widths and 2-star hyperon resonances. We present new results on  strangeness yield as a function of centrality and present their interpretation in terms of QGP inspired model of strangeness abundance in the hadronizing fireball.
\end{abstract}

\section{Inclusion of new data}
We interpret strange soft hadron  multiplicity results obtained in Pb--Pb collisions at $\sqrt{s_{NN}}=2.76\,\mathrm{TeV}$ at CERN Large Hadron Collider (LHC). This  contribution extends the more  detailed presentation of Ref.~\cite{Petran:2013lja}, whereas  Ref.~\cite{Petran:2013qla}  provided the related analysis without data inter- or extrapolation. In the results presented here we can use final data for K$_S^0$, $\Lambda$, $\Xi^\pm$ and $\Omega^\pm$~\cite{Abelev:2013xaa,ABELEV:2013zaa}. Our  fit with these results converges to the same set of thermal parameters as in~\cite{Petran:2013lja}. 

\begin{figure}[t]
\begin{center}
\includegraphics[width=0.62\columnwidth]{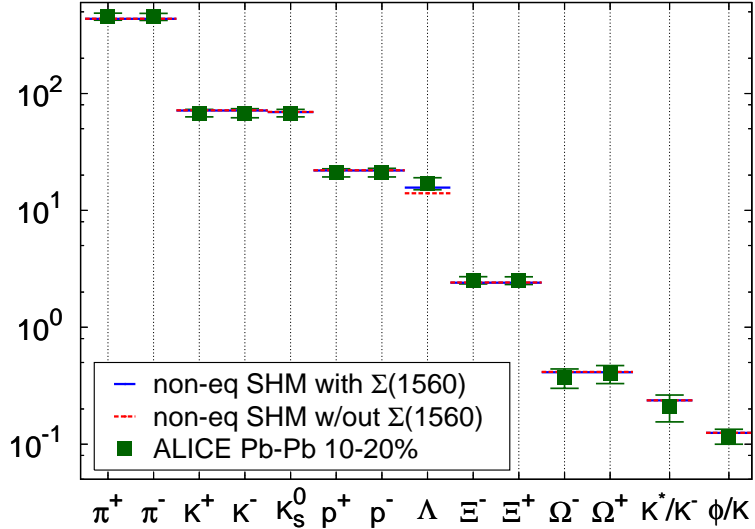}
\caption{\label{fig:fittedparticles}(color online) Comparison of non-equilibrium SHM fit with most recent data from Pb--Pb collisions at $\sqrt{s_{NN}}=2.76\,\mathrm{TeV}$ for the 10\%--20\% centrality bin measured by the ALICE collaboration (darkgreen squares), with the 2-star  resonance $\Sigma(1560)$ feeding the yield of $\Lambda$ (solid blue bars), and without the $\Sigma(1560)$ feed (red dashed bars).}
\end{center}
\end{figure}

Considering limited space and in order to rely solely on directly measured and final experimental results we show here as an example the centrality 10\%--20\% data bin. In the top two result rows of \Tref{tab:thermalparameters}, we compare the prior with the present results for statistical parameters. As before  $\Lambda$-abundance is the dominant contributor to the error as is seen also in direct comparison of input and output results shown in \Fref{fig:fittedparticles}. Thus in \Tref{tab:thermalparameters} the $\chi^2_\mathrm{tot}$ is found to be  greater, yet relatively small given the 9 degrees of freedom. 

\begin{table}[b]
\caption{\label{tab:thermalparameters}Comparison of  non-equilibrium SHM model parameter obtained in fit to   10--20\% centrality  results for Pb--Pb collisions at $\sqrt{s_{NN}}=2.76\,\mathrm{TeV}$. See text for discussion. }
\lineup
\resizebox{\columnwidth}{!}{
\begin{tabular}{*{8}{c}}
\br
$\Sigma(1560)$ & widths & $dV/dy\,[\mathrm{fm}^3]$ & $T\,[\mathrm{MeV}]$& $\mu_B\,[\mathrm{MeV}]$ & $\gamma_q$ & $\gamma_s$ & $\chi^2_\mathrm{tot}$ \\
\mr
\multicolumn{2}{l}{reported in~\cite{Petran:2013lja}}   
          & $2003\pm\ 47$ & $138.6$   & $1.23\pm0.06$  & $1.63\pm0.01$ & $2.06\pm0.13$ & $3.943$\\
NO  & NO  & $2033\pm105$ & $138.6$   & $1.36\pm7.94$  & $1.63\pm0.01$ & $2.02\pm0.08$ & $6.599$\\
NO  & YES & $1978\pm488$ & $139.3$   & $1.15\pm0.97$  & $1.62\pm0.19$ & $2.00\pm0.31$ & $5.169$\\
YES & NO  & $2042\pm409$ & $138.5$   & $0.49\pm0.35$  & $1.63\pm0.15$ & $2.03\pm0.25$ & $4.766$\\
YES & YES & $1976\pm398$ & $139.2$   & $0.74\pm0.09$  & $1.62\pm0.01$ & $2.01\pm0.14$ & $3.472$\\
\br
\end{tabular}
}
\end{table}

As seen in the top two rows of results, the differences between these two fits  are well within the error. Looking at $\mu_B$, recall that we did constrain the value in Ref.~\cite{Petran:2013lja} by its centrality dependence. Our present result is the outcome of an unconstrained fit for this centrality.  This is possible since the added experimental data for  $\Xi^\pm$ and $\Omega^\pm$ contain a small particle-antiparticle asymmetry even if this is well within the particle yield error bar. The value  of $\mu_B$  we find is near to our earlier expectations; however, the error in $\mu_B$ is  large. There is no error shown for $T$ since the statistical fit error is much smaller than the precision of the result shown. The error that relates directly to the error of fitted particle yields is found to mainly impact $dV/dy$ and $\gamma_s$. This shows that there are two types of error in the data: a common normalization error for all data points, and a further strange particle normalization error. 

The bulk properties of the fireball at hadronization are shown in \Tref{tab:physicalproperties}.  These are presented without an error estimate: the impact of the error in $dV/dy$  impacts directly  the  total strangeness yield  $N_{s+\bar{s}}$ and  entropy $S$. The other three properties that we show, pressure $P$, energy density $\varepsilon$, trace anomaly $(\varepsilon-3P)/T^4$, are  impacted by other statistical parameters but the error of $\gamma_s$ is the most important. The two top rows are practically equal showing that the additional data, and changes in final data do not impact any of the results and discussion presented in Ref.~\cite{Petran:2013lja}.

\begin{table}[b]
\caption{\label{tab:physicalproperties} Comparison of models (see table \ref{tab:thermalparameters} for all other details) in terms of physical bulk properties. From left to right, we show pressure $P$, energy density $\varepsilon$, trace anomaly $\varepsilon-3P$ in units of $T^4$, total strangeness $N_{s+\bar{s}}$ and entropy $S$.}
\lineup
\begin{center}
\begin{tabular}{*{7}{c}}
\br
$\Sigma(1560)$ & widths & $P\,[\mathrm{MeV/fm}^3]$ & $\varepsilon\,[\mathrm{GeV/fm}^3]$ & $(\varepsilon - 3P)/T^4$ & $N_{s+\bar{s}}$ & $S$  \\
\mr
\multicolumn{2}{l}{reported in~\cite{Petran:2013lja}}   
             & $79.1$ & $0.467$   & $4.77$  & $384$ & $6466$ \\
NO  & NO     & $79.2$ & $0.468$   & $4.79$  & $388$ & $6589$ \\
NO  & YES    & $82.0$ & $0.483$   & $4.85$  & $385$ & $6593$ \\
YES & NO     & $78.7$ & $0.465$   & $4.78$  & $387$ & $6574$ \\
YES & YES    & $81.7$ & $0.481$   & $4.84$  & $384$ & $6570$ \\
\br
\end{tabular}
\end{center}
\end{table}

\section{Improving the fit}
\subsection{Influence of hadron resonance width}
The SHARE package~\cite{Torrieri:2004zz,Torrieri:2006xi} was developed with the option of allowing  the   resonance widths presented by the Particle Data Group~\cite{Beringer:1900zz} (PDG) values. The computing time required to allow the resonance widths is significant with fits needing often more than 24h CPU time. The reason for this is that many numerical foldings need to be performed in order to evaluate the total particle yields; see~\cite{Torrieri:2004zz} for  details. In the past the small change in outcome did not justify the numerical effort. Thus this test has not been performed often.  

The third result row in \Tref{tab:thermalparameters} shows the statistical parameters obtained allowing for hadron resonance widths. While $\chi^2_\mathrm{tot}$ of the fit slightly decreases and central points of statistical parameters aside of $T$ agree within error, we note that the error is now more distributed among parameters and the progression  is larger indicating slightly less consistency  between fit and data. 

The fact that there is an increase in $T$ by $\delta T=0.7$ MeV, which is well above statistical error, is not surprising considering  the systematic effect that the resonance widths have on particle yields. Note that to compensate the increase in value of $T$ the central value of $dV/dy$ decreased. Another consequence of the slight increase of $T$ is a slight  (1\%) increase of the intensive bulk properties  $P$,  $\varepsilon$,   $(\varepsilon-3P)/T^4$, see the third result row in \Tref{tab:physicalproperties}. The extensive bulk properties directly related to the measured particle yields  i.e.  $N_{s+\bar{s}}$ and  $S$ remain unchanged.

\subsection{$\Sigma(1560)$ as a source of $\Lambda$}
The only data point that is not fitted within the 1 s.d. experimental error margin is the yield of $\Lambda$. This situation was also discussed  in Ref.~\cite{Petran:2013lja}, where the preliminary $\Lambda/\pi$ ratio was fitted and was systematically under--predicted for all centralities as the only data point standing out with model value just outside the one standard deviation error margin. The newly reported experimental value for 10--20\% centrality, $\Lambda=17\pm2$, is to be compared to the fitted value,  $\Lambda=14$. Inclusion of hadron resonances with their widths discussed above does not   improve  the fit to the $\Lambda$ data point. 

The question arises if perhaps a contribution to the  $\Lambda$ yield was inadvertently omitted. The SHARE implementation of SHM~\cite{Torrieri:2004zz,Torrieri:2006xi} includes  3-star (***) and 4-star (****) hadron resonances from among all resonances reported by PDG~\cite{Beringer:1900zz}. There are many omitted 2-star (**) resonances. Generally  these are so massive that their contribution to particle yields is insignificant. However,  we found one exception to this rule, a $\Sigma(1560)$ which decays into $\Lambda$ and which would thus systematically increase the $\Lambda$ yield within the error across all centrality.

The charged states $\Sigma(1560)^\pm$ have been clearly observed ($6\sigma$ signal) by two independent experiments~\cite{Meadows:1980vr}. On the other hand, no evidence of $\Sigma(1560)^0 \to \Lambda\pi^0$ has been found in a recent crystal ball experiment~\cite{Olmsted:2003is} and hence the resonance remains unconfirmed at (**) level.  $\Sigma(1560)$ quantum numbers have not been measured. Inspired by the close in mass resonance $\Sigma(1580)\frac{3}{2}^-$~\cite{Litchfield:1974ru}, we assign spin $\frac{3}{2}$ also to $\Sigma(1560)$. 

Once we include $\Sigma(1560)$ in the list of hadron states, the new resonance decay feeds the $\Lambda$ yield by $\Sigma(1560)\to\Lambda\pi$ (100\%) decay. When we repeat the fit, the resulting statistical model parameters are all almost identical to those without $\Sigma(1560)$  as is seen in \Tref{tab:thermalparameters}. This means that the other 13 data points alone constrain enough the fit parameters so that the effect of an omitted resonance simply reduces the $\chi^2$ without changing other results. This result confirms `missing' resonance hypothesis as the probable origin of the $\Lambda$ yield underprediction.

In the final result row in \Tref{tab:thermalparameters}, we consider the finite width of all resonances and the influence of the 2-star $\Sigma(1560)$ resonance. The overall $\chi^2_\mathrm{tot}$ is decreased by a factor of two,  the fitted value of $\mu_B$ is showing a relatively small error. The main fit error is in overall normalization $dV/dy$. The physical bulk properties remain as already obtained and discussed without $\Sigma(1560)$.

\section{Strangeness conservation and strange quark mass at hadronization}
\begin{figure}[t]
\centering
\begin{minipage}[t]{0.48\columnwidth}
\includegraphics[height=50mm]{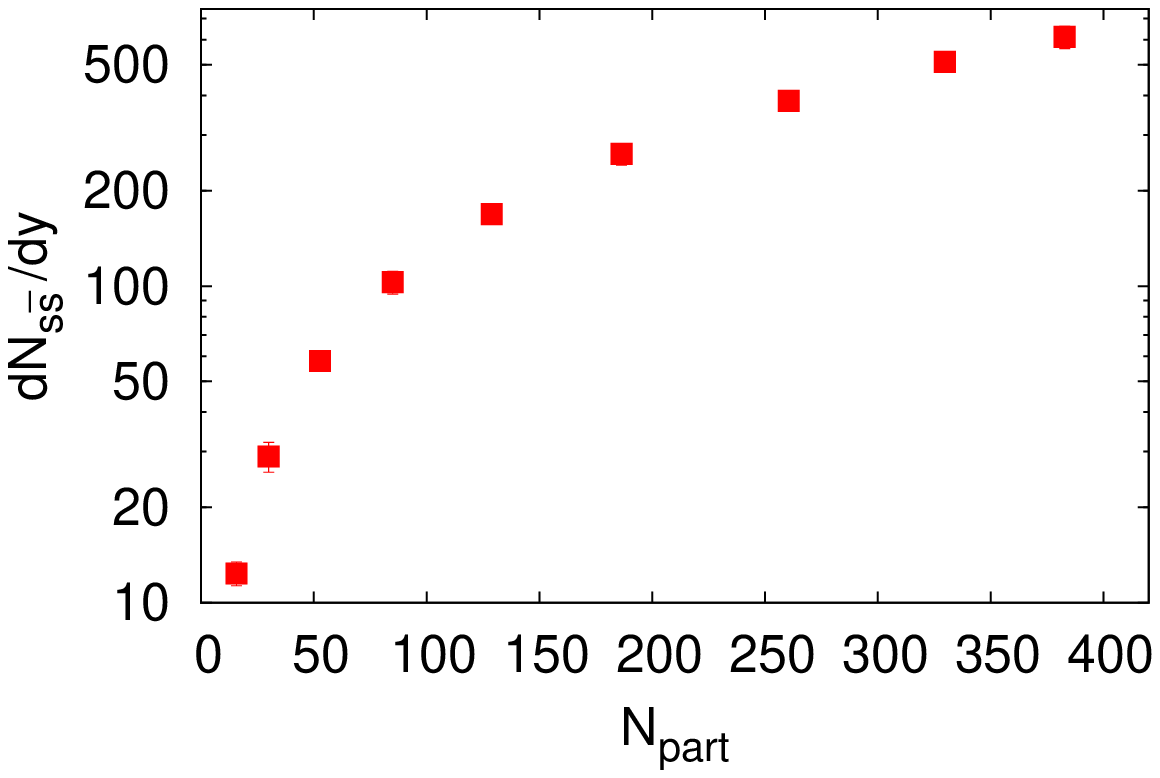}
\caption{\label{fig:strangenessNumber}(color online) Total strangeness $dN_{s+\bar{s}}/dy$ produced in Pb--Pb collisions at $\sqrt{s_{NN}}=2.76\,\mathrm{TeV}$ as a function of centrality.}
\end{minipage}\hspace*{0.04\columnwidth}%
\begin{minipage}[t]{0.48\columnwidth}
\includegraphics[height=50mm]{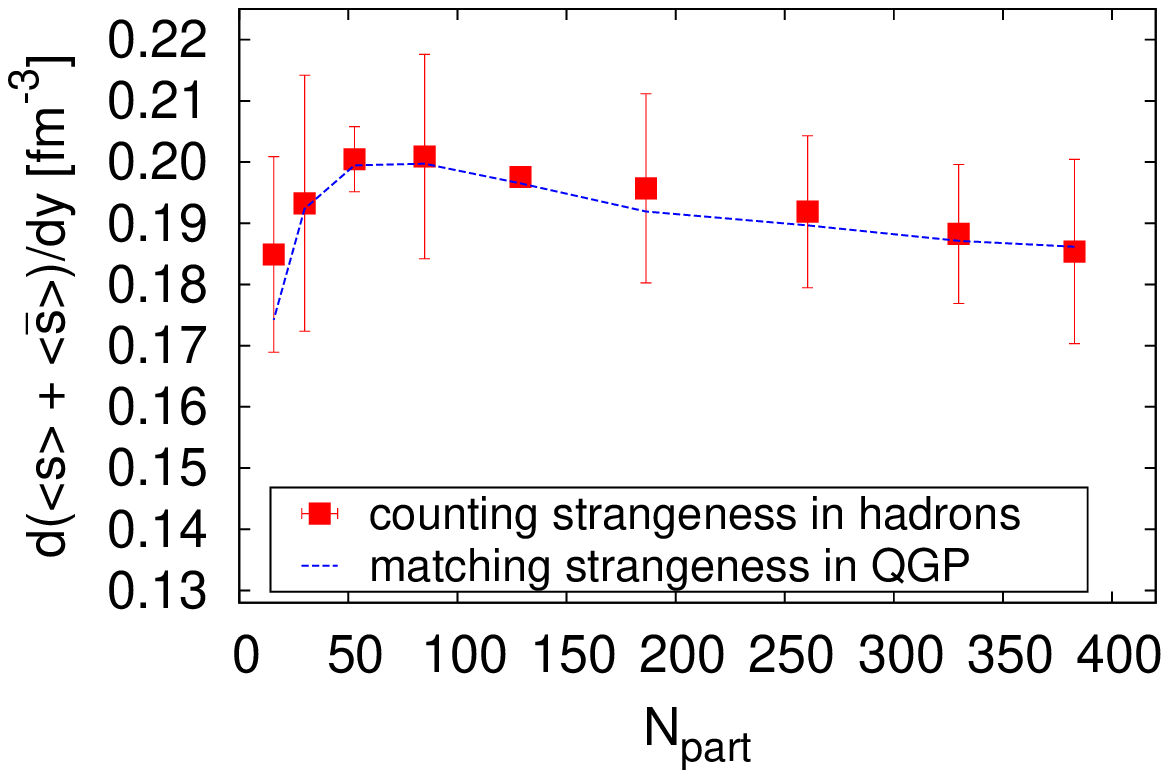}
\caption{\label{fig:strangeness}(color online) $s+\bar s$ strangeness density per unit rapidity in the hadron phase (red squares), fitted with strangeness in the QGP phase (blue dashed line). See text.}
\end{minipage}
\end{figure}

In the central Pb--Pb collisions at the LHC, an unprecedented amount of strangeness is produced, a total of $dN_{s\bar{s}}/dy\simeq 600$ strange and anti-strange quarks per unit of rapidity for the most head-on  5\% central collisions, see \Fref{fig:strangenessNumber}. For the peripheral collisions the rise of the total strangeness yield is very rapid, as both the size of the volume and saturation of strangeness production combine. For the more head-on collisions we see a power ($\sim1.17$) law rise similar to particle interpolation in~\cite{Petran:2013lja}. Normalization error bar contained in $dV/dy$ is typically 10\% is not shown, only the   error from  $\gamma_s$ is shown in \Fref{fig:strangenessNumber} (often hidden in the symbol size).

\begin{figure}[t]
\centering
\begin{minipage}[t]{0.48\columnwidth}
\includegraphics[height=51mm]{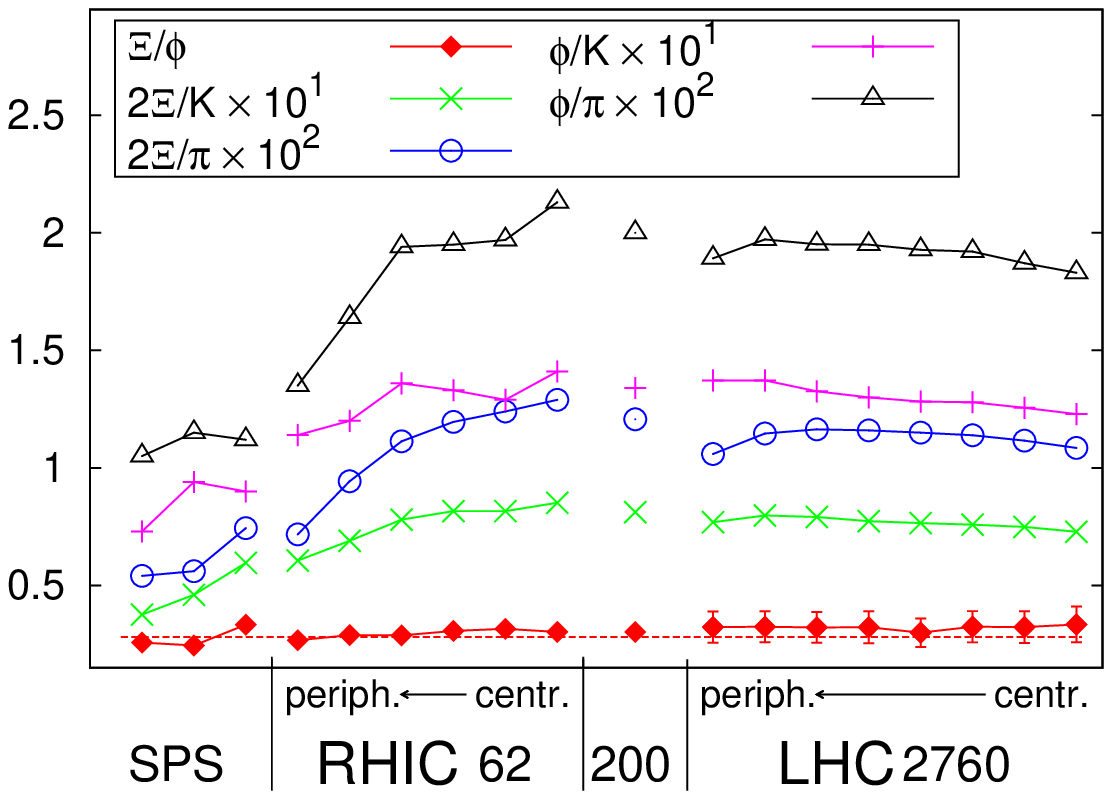}
\caption{\label{fig:ratios} Measured particle ratios from relativistic heavy ion collisions across different collisional energies at SPS, RHIC and LHC, and centralities.}
\end{minipage}\hspace*{0.04\columnwidth}%
\begin{minipage}[t]{0.48\columnwidth}
\includegraphics[height=50mm]{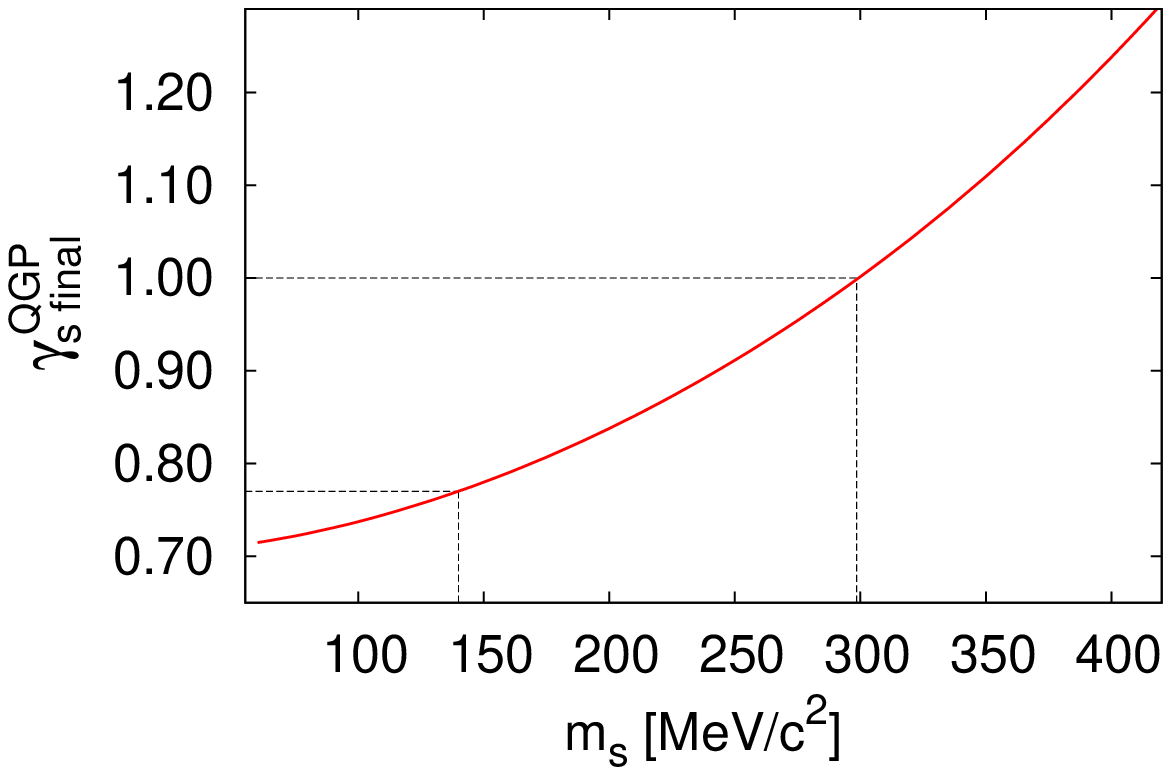}
\caption{\label{fig:gammasQGP}(color online) The strangeness phase space occupancy in the QGP phase as defined by Eq.~\eref{eq:gammasQGP} and correlated in the fit to strangeness density with the strange quark mass. See text.}
\end{minipage}
\end{figure}

The corresponding total observed strangeness density in the hadron phase is depicted in \Fref{fig:strangeness}. The uncertainty is evaluated using the relative uncertainty of the model parameter $\gamma_s$ reported in~\cite{Petran:2013lja}. Note that the variation in density shown is only about $\pm3.5$\%. Such constancy of the strangeness yield, assuming a recombinant hadronization, should result in strange hadron yield ratios independent of centrality.  In \Fref{fig:ratios} the flat line at the bottom is a ratio of two doubly strange particle yields, $\Xi/\phi$. Its constancy provides a reference of precision of the argument as this result should be constant~\cite{Petran:2009dc} even if the strangeness density varies.

The other ratios compare the yield of doubly strange particles with the yield of single strange particles and in one case doubly strange with pions. When strangeness is not saturated as a function of centrality in QGP, one expects an increase in these ratios which we clearly see at lower SPS and RHIC energies. However at LHC a different pattern emerges: there is a bit of increase looking at some of the most peripheral bins which is followed by a slow decrease. The strangeness density shown in \Fref{fig:strangeness} mirrors this behavior.

It is of considerable interest to understand if the `measured' strangeness density seen in  \Fref{fig:strangeness} can be interpreted in terms of sudden hadronization of a QGP fireball. Sudden hadronization implies the conservation of strangeness yield. In sudden hadronization model the volume does  not change  thus the density in the hadron phase equals that in  QGP as well.   The chemical freeze-out temperature $T$  is also the hadronization temperature, i.e. the  temperature of QGP breakup. We evaluate the QGP  phase strangeness density for a given $T$ given by integral of Fermi gas strangeness density in the QGP  using~\cite{Letessier:2002gp}:
\begin{equation}
\label{eq:strangeness}
s(m_s,T; \gamma_s^{\scriptscriptstyle\mathrm{QGP}}) = -\frac{g}{2\pi^2} \left(\frac{T}{\hbar c} \right)^3 
\sum\limits_{n=1}^\infty \left(-\gamma_s^{\scriptscriptstyle\mathrm{QGP}}\right)^{n} \frac{1}{n^3} \left(\frac{nm_s}{T} \right)^2 K_2\left(\frac{nm_s}{T} \right),
\end{equation}
where  $m_s$ is the strange quark mass, $\gamma_s^{\scriptscriptstyle\mathrm{QGP}}$ is the phase space occupancy: here  superscript  QGP helps to distinguish  from that in the hadron phase  where we used $\gamma_s$ without a superscript. The degeneracy $g=12=2_\mathrm{spin} 3_\mathrm{color}  2_\mathrm{p}$  where the last factor accounts for the presence of both quarks and antiquarks. We will discuss below the reduction in $s$ due to color-interactions.

In central LHC  collisions, the large volume (longer lifespan) suggests that strangeness approaches saturated yield  in the QGP. This hypothesis is confirmed  by almost constant particle ratios we discussed above, see  \Fref{fig:ratios}. However, in peripheral collisions, the short lifespan of the fireball may not be sufficient to reach chemical equilibrium. Therefore we introduce a centrality dependent strangeness phase space occupancy $\gamma_s^{\scriptscriptstyle\mathrm{QGP}}(N_{part})$ which is to be used in Eq.\eref{eq:strangeness}. 

To model the centrality dependence of $\gamma_s^{\scriptscriptstyle\mathrm{QGP}}(N_{part})$, we recall that  the lifespan $\tau$ of the fireball  depends on the transverse surface of the fireball and hence the time is proportional to the transverse radius $\tau\propto r_\perp$. The transverse radius is related to number of participants as $r_\perp \propto N_{part}^{1/3}$. Thus, at mid-rapidity, the strangeness production will be assumed to be proportional to $r_\perp^2 \propto N_{part}^{2/3}$ and modeled by the usual saturating functional form:
\begin{equation}
\label{eq:gammasQGP}
\gamma_s^{\scriptscriptstyle\mathrm{QGP}}(N_{part}) = \gamma_{s\,\mathrm{final}}^{\scriptscriptstyle\mathrm{QGP}}\tanh\left[\left(\frac{N_{part}}{N_0}\right)^{^2\!/\!\!\;_3}\right].
\end{equation}
$\gamma_{s\,\mathrm{final}}^{\scriptscriptstyle\mathrm{QGP}}$  is the asymptotic saturation  of strangeness phase space in the QGP for large systems and $N_0$ controls the scale of the fireball transverse size.

We now match the strangeness density measured in the hadron phase shown in \Fref{fig:strangeness} with the centrality dependent strangeness density in the QGP, using Eq.\eref{eq:gammasQGP}  in Eq.\eref{eq:strangeness}. Our fit to the observed strangeness density is shown in \Fref{fig:strangeness}. We observe a very strong correlation between $m_s$ and $\gamma_{s\,\mathrm{final}}^{\scriptscriptstyle\mathrm{QGP}}$, which we show as a function in \Fref{fig:gammasQGP}, while  choosing the best $N_0$ which converges within a narrow interval of $N_0 \in (11, 14)$. This shows a quick saturation of $\gamma_s^{\scriptscriptstyle\mathrm{QGP}}(N_{part})$, which reaches $0.95\,\gamma_{s\,\mathrm{final}}^{\scriptscriptstyle\mathrm{QGP}}$ already for $N_{part}\simeq 2 N_0 = 30$. 

To choose a set of values of ($m_s$, $\gamma_{s\,\mathrm{final}}^{\scriptscriptstyle\mathrm{QGP}}$) we consider two cases shown in \Fref{fig:gammasQGP}:
\begin{enumerate}
\item The strangeness in QGP is chemically equilibrated in central collisions, $\gamma_{s\,\mathrm{final}}^{\scriptscriptstyle\mathrm{QGP}}\simeq 1$. This requires the strange quark to have an effective mass of $m_s=299\,\mathrm{MeV}/c^2$ at hadronization;
\item We assume the PDG value of strange quark mass~\cite{Beringer:1900zz} $m_s \simeq 140\,\mathrm{MeV}/c^2$  at a scale of $\mu\simeq 2\pi T \simeq 0.9\,\mathrm{GeV}$. This requires $\gamma_{s\,\mathrm{final}}^{\scriptscriptstyle\mathrm{QGP}}\simeq 0.77$. 
\end{enumerate}
We believe  that both approaches can coincide in an improved theoretical description: 
\begin{enumerate}
\item We omitted in Eq.~\eref{eq:strangeness} the  thermal QCD prefactor which reduces the expected QGP strangeness density due to thermal QCD many body interactions. The  typical reduction of $s$ is by a factor   $(1-c \alpha_s/\pi)$. With  $\alpha_s\simeq 0.65$  and $c\simeq 1$ this effect reduces the QGP density by $\simeq 20$\%. The resultant $\gamma_{s\,\mathrm{final}}^{\scriptscriptstyle\mathrm{QGP}}$ can be 20\% larger. This effect is present; we do not know its exact magnitude. 
\item The measured value of $s$  could be reduced from hadronization value by longitudinal dilution of strangeness during matter expansion after hadronization. This effect vanishes in the limit of Bjorken scaling as for every particle that moves out, another particle moves back into the central rapidity acceptance domain. The rapidity plateau has not been demonstrated experimentally for strange hadrons at LHC. 
\end{enumerate}
We note that longitudinal dilution by  15\%   restores strangeness abundance to prior expectations~\cite{Rafelski:2010cw}. Both effects would allow $m_s \simeq 140\,\mathrm{MeV}/c^2$ to be consistent with $\gamma_{s\,\mathrm{final}}^{\scriptscriptstyle\mathrm{QGP}}\simeq 1$.

\section{Conclusions}
The most important result of this analysis is complete stability at 1\%-level of results presented in Ref.~\cite{Petran:2013lja}. This earlier analysis is fully compatible with the latest  results~\cite{Abelev:2013xaa,ABELEV:2013zaa}. The 1--1.5 s.d. discrepancy of $\Lambda$ yield systematically below the experimental result inspired us to explore potential  $\Lambda$ sources. We investigated the finite width of all resonances. We found that  our fit is very stable and confidence level is slightly improved, however to explain  $\Lambda$ yield we needed  the  2-star (**) resonance $\Sigma(1560)$, increasing the model yield of $\Lambda$ to within $^1\!/\!\!\:_2$ s.d. of the experimental yield. $\Sigma(1560)$ causes no other change in the outcome of our analysis.

We than considered in detail how to interpret the strangeness yield present at hadronization in terms of a QGP inspired model. We considered   strangeness conservation during hadronization and concluded that for a fully consistent description we must account for possible reduction of $s$ by interactions. We than argued that the relatively low hadronization strangeness density could be a consequence of particle dilution in central rapidity region should Bjorken scaling not apply fully. Hadrons from jet quenching and charm decay are produced predominantly in central rapidity domain.

\section*{\label{sec:acknowledgements}Acknowledgments}
This work has been supported by a grant from the U.S. Department of Energy, grant DE-FG02-04ER41318, Laboratoire de Physique Th{\' e}orique et Hautes Energies, LPTHE, at University Paris 6 is supported by CNRS as Unit{\' e} Mixte de Recherche, UMR7589. 

\end{document}